\documentclass[aps,twocolumn,pra]{revtex4}
\usepackage{amsmath,amssymb,bm}
\usepackage{graphicx}
\usepackage{epstopdf}
\usepackage{latexsym}
\usepackage{subfigure}
\usepackage[usenames,dvipsnames]{color}
\usepackage{hyperref}
\usepackage{natbib}
\usepackage{moreverb}

\begin{document}
\newcommand{\MF}{CH$_3$F}

\newcommand{\todo}[1]{{\color{red}#1}}

\newcommand{\tj}[6]{\left(\begin{array}{ccc} #1&#3&#5\\ #2&#4&#6\end{array}\right)}
\newcommand{\sj}[6]{\left\{\begin{array}{ccc} #1&#3&#5\\ #2&#4&#6\end{array}\right\}}

\title{Realizing unconventional quantum magnetism with symmetric top molecules }

\author{M. L. Wall$^{1,2}$\footnote{e-mail: mwall.physics@gmail.com},  K. Maeda$^{2}$, and L. D. Carr$^{2}$}

\affiliation{$^{1}$JILA, NIST, Department of Physics, University of Colorado, Boulder, CO 80309, USA}
\affiliation{$^{2}$Department of Physics, Colorado School of Mines, Golden, Colorado 80401, USA}

\begin{abstract}

We demonstrate that ultracold symmetric top molecules loaded into an optical lattice can realize highly tunable and unconventional models of quantum magnetism, such as an XYZ Heisenberg spin model.  We show that anisotropic dipole-dipole interactions between molecules can lead to effective spin-spin interactions which exchange spin and orbital angular momentum.  This exchange produces effective spin models which do not conserve magnetization and feature tunable degrees of spatial and spin-coupling anisotropy.  In addition to deriving pure spin models when molecules are pinned in a deep optical lattice, we show that models of itinerant magnetism are possible when molecules can tunnel through the lattice.  Additionally, we demonstrate rich tunability of the effective models' parameters using only a single microwave frequency, in contrast to proposals with $^1\Sigma$ diatomic molecules, which often require many microwave frequencies.  Our results are germane not only for experiments with polyatomic symmetric top molecules, such as methyl fluoride (\MF), but also diatomic molecules with an effective symmetric top structure, such as the hydroxyl radical OH.

\end{abstract}

\maketitle

\section{Introduction}

Lattice models of exchange-coupled quantum mechanical spins such as the Heisenberg model have long served as paradigmatic examples of strongly correlated many-body systems~\cite{Auerbach,SchollwockBook}.  The exquisite tunability and precise microscopic characterization of ultracold gases makes them promising candidates for exploring quantum magnetism.  However, the most prominent platform for ultracold gas quantum simulation, neutral atoms loaded into optical lattices, has difficulty reaching the regime where quantum magnetism is manifest~\cite{WHR_Review,lewensteinM2007,bloch2008}.  The reason for the difficulty is that the short-range interactions experienced by neutral atoms require two atoms to occupy the same lattice site in order to significantly interact.  For two-component (effective spin-1/2) atoms, effective models of quantum magnetism emerge when on-site interactions $U$ are significantly larger than the tunneling amplitude $t$ between neighboring lattice sites, pinning the atoms in a Mott insulator phase with one atom in each lattice site.  Effective spin interactions are then mediated by a superexchange process~\cite{Auerbach} which requires virtual tunneling to doubly occupied sites.  Because the resulting effective spin couplings scale as $t^2/U$ with $t\ll U$, the temperature scales required to see the onset of magnetism are extraordinarily small.

Systems which feature long-range interactions can generate effective spin-spin interactions which are not mediated by tunneling, and so can display coherent internal state many-body dynamics even without quantum degeneracy in the motional degrees of freedom.  Such long range effective spin couplings have been realized using trapped ions~\cite{PhysRevLett.92.207901,kim2010quantum,britton2012engineered}, Rydberg atoms~\cite{Weidemueller}, and magnetic atoms~\cite{Laburthe}, and have been proposed for other platforms, such as atoms in optical cavities~\cite{PhysRevLett.95.260401}.  In this work, we focus on the realization of long-range effective spin interactions with polar molecules, as has been recently demonstrated experimentally~\cite{Yan_Moses, Hazzard_Gadway_14}.  A unique feature of dipolar realizations of quantum magnetism, as polar molecules in optical lattices provide, compared to non-dipolar systems (e.g.,~trapped ions) is that dipolar interactions are anisotropic.  Anisotropic interactions do not conserve the internal (e.g.,~rotational) or the spatial angular momentum separately, but only their sum.  By mapping the internal angular momentum of a molecule, in particular its rotational angular momentum, to an effective spin, the dipole-dipole interaction hence generates the possibility of unconventional models of quantum magnetism which do not conserve the total magnetization.  As we will show in this paper, such models feature tunable degrees of both spin and spatial anisotropy.

\begin{figure}[tbp]
\centerline{\includegraphics[width=0.8 \columnwidth]{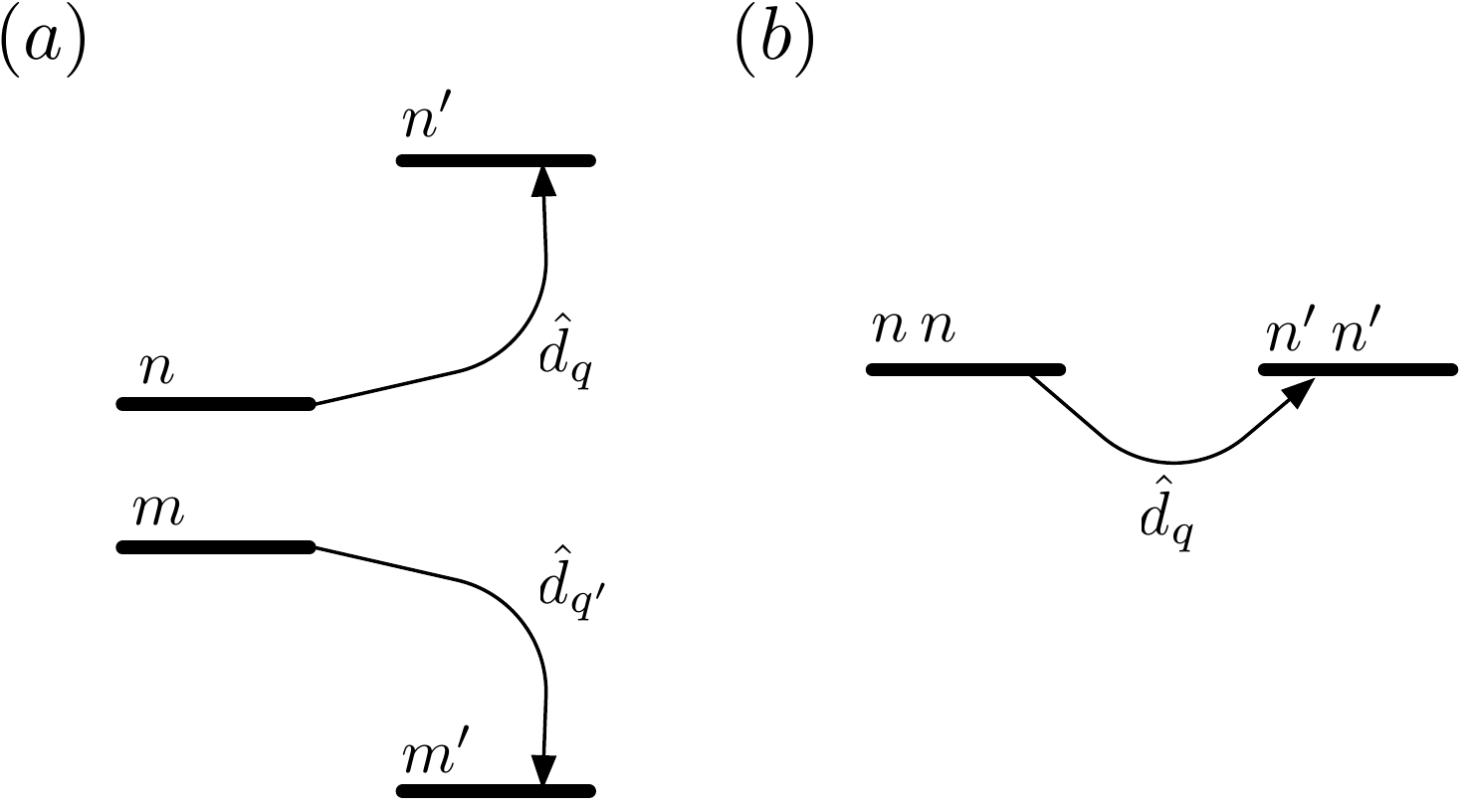}}
\caption{\label{fig:Resonance} 
\emph{{Resonant dipolar processes which exchange internal and external angular momentum}} (a) An example of a F\"orster resonance which involves 4 different internal states satisfying the resonance condition $E_n+E_m\simeq E_{n'}+E_{m'}$.  Single-particle dipole-allowed transitions $|n\rangle\to|n'\rangle$ and $|m\rangle\to |m'\rangle$ drive the interaction-induced two-particle transition $|nm\rangle\to|n'm'\rangle$.  (b) The resonances utilized in this work involve single-particle levels $|n\rangle$ and $|n'\rangle$ which are nearly degenerate and posses a single-particle dipole-allowed transition $|n\rangle\to|n'\rangle$ that may vanish upon time averaging.  Interactions cause a two-particle transition $|nn\rangle\to |n'n'\rangle$, changing the net rotational projection of the molecules.}
\end{figure}

The exchange of internal and external angular momentum projection by dipole-dipole interactions requires two pairs of internal states which are nearly degenerate in energy (on the scale of dipole-dipole interactions) and also have dipole-allowed transitions between them.  We show two such scenarios in Fig.~\ref{fig:Resonance}.  The first scenario is that we have two pairs of internal states, call them $(n,m)$ and $(n',m')$ with energies $E_{n}+E_{m}\simeq E_{n'}+E_{m'}$ nearly degenerate.  Further, we assume that at least one of the latter states is not a member of the former pair of states~\footnote{Processes which exchange two internal states through the transition $|n m\rangle\to|m n\rangle$ do not change the total angular momentum projection by definition, and so are not of the type we are discussing here.  However, these terms do contribute to the total Hamiltonian, see, e.g.,Eq.\eqref{eq:SpinEx}.}, see Fig.~\ref{fig:Resonance}(a).  Such a two-particle near-degeneracy with non-radiative dipole coupling is generally called a F\"orster resonance, and such resonances have been fruitfully applied to control the interactions in Rydberg atoms~\cite{PhysRevLett.97.083003,PhysRevLett.108.113001}.  Additionally, such resonances may occur at isolated points in the spectra of $^1\Sigma$ polar molecules, those with no orbital or spin angular momentum~\cite{Chiron,Gorshkov_Manmana_11b}.

\begin{figure}[bp]
\centerline{\includegraphics[width=\columnwidth]{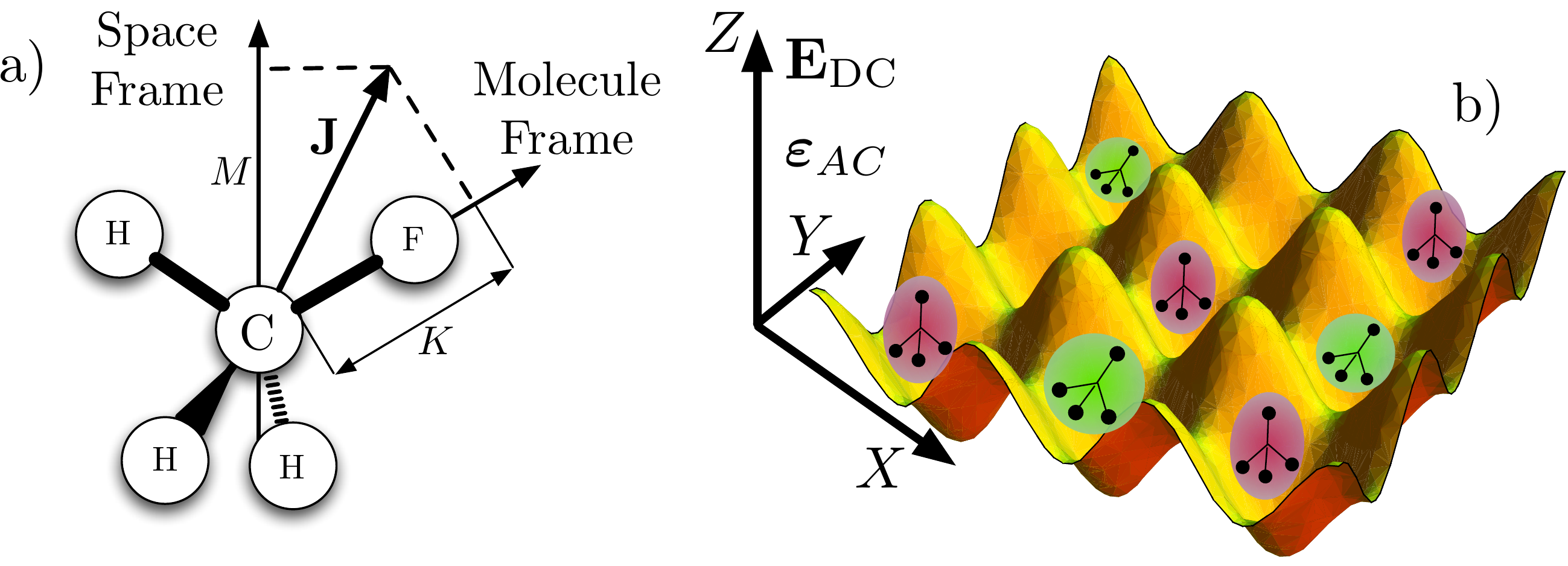}}
\caption{\label{fig:Levels} 
\emph{Symmetric top molecules (STMs) in optical lattices} (a) Rotational angular momentum geometry of the STM {\MF}.  (b) Schematic of field and lattice geometry.  Purple and green denote two internal states.}
\end{figure}

In this work, we instead exploit a resonant process such as is shown in Fig.~\ref{fig:Resonance}(b).  Here, two particles in the same internal state $n$ are transferred to a different internal state $n'$ which is dipole-coupled to the first and brought into resonance by external fields.  In contrast to the F\"orster resonance, this latter type of resonance involves only two single-particle states, and so naturally leads to a description in terms of a spin-1/2 system.  Such resonances are a generic feature of magnetic dipoles, and lead to phenomena such as spontaneous demagnetization of spinor Bose gases~\cite{PhysRevLett.106.255303}.  In contrast, electric dipoles have parity and time-reversal selection rules which would appear to preclude a dipole-coupled resonance such as is shown in Fig.~\ref{fig:Resonance}(b).  A key finding of this work is that resonances like Fig.~\ref{fig:Resonance}(b) with an electric dipole transition are also a generic feature of symmetric top molecules (STMs) with microwave and static field dressing, even in the presence of hyperfine or other detailed molecular structure.  Polyatomic STMs, such as the methyl fluoride molecule, {\MF}, shown in Fig.~\ref{fig:Levels}(a), have a high degree of symmetry in their rotational structure which makes them behave as ``electric analogs" of pure magnetic dipoles~\cite{Wall_Maeda_13}.  Further details on STMs and their interactions with external fields is provided in Sec.~\ref{sec:STM}.

In the present work, we show that two isolated internal states of an STM tuned near a resonance of the form shown in Fig.~\ref{fig:Resonance}(b) form an effective spin-1/2 which is governed by a model with tunable anisotropy in both the spatial and spin-component dependence of the effective spin couplings.  In contrast to related proposals, such as the realization of spin-component anisotropic XYZ Heisenberg models using bosonic atoms in excited $p$-orbitals of an optical lattice~\cite{PhysRevLett.111.205302} or in a synthetic gauge field~\cite{SOC1,SOC2}, our spin couplings are non-perturbative in the particle-particle interaction strength and are not mediated by tunneling through the lattice.  Hence, magnetic phenomena in our proposal can be realized even in the absence of motional quantum degeneracy for the molecules.  The ability to observe coherent many-body dynamics in a non-degenerate sample of ultracold molecules is important, as cooling of molecules is difficult and the fully quantum degenerate regime has not yet been reached~\cite{WHR_Review,Quemener_Julienne_Review}.  Additionally, our microwave dressing proposal applies for present $\sim 500$nm optical lattice spacings, in distinction to proposals for $^2\Sigma$ molecules, where appreciable couplings require trapping at quite small lattice spacings~\cite{Brennen2}.  Finally, our proposal requires only a single microwave frequency, in contrast to many proposals for $^1\Sigma$ molecules where multiple frequencies are required~\cite{Gorshkov_Manmana_11,Gorshkov_Manmana_11b,PhysRevB.87.081106,doi:10.1080/00268976.2013.800604}.

This paper is organized as follows.  Sec.~\ref{sec:XYZ} provides a phenomenological analysis of how the resonances shown in Fig.~\ref{fig:Resonance}(b) lead to effective quantum spin systems, e.g.~XYZ Heisenberg spin models, which do not conserve magnetization.  In Sec.~\ref{sec:STM} we present an overview of symmetric top molecules and their coupling to external fields.  In particular, Sec.~\ref{sec:Microwave} discusses the interaction of symmetric top molecules with microwave radiation that is near-resonant with a rotational transition.  We then focus on how to engineer the external fields to obtain tunable resonances like those in Fig.~\ref{fig:Resonance}(b), and analyze the dipole-dipole interactions between microwave-dressed states.  Section \ref{sec:EffectiveModels} derives effective many-body models of quantum magnetism which are applicable near the field-induced level crossings.  Finally, in Sec.~\ref{sec:outlook}, we conclude.  Some more technical details of the dipole-dipole interactions in the microwave-dressed basis states are given in the appendix.

\section{Phenomenological realization of an XYZ spin model}
\label{sec:XYZ}
In this section, we provide a phenomenological analysis of how a resonance such as that shown in Fig.~\ref{fig:Resonance}(b) leads to an effective XYZ spin model for a collection of STMs pinned in a quasi-2D lattice geometry, as in Fig.~\ref{fig:Levels}(b).  A more detailed analysis will be provided in Sec.~\ref{sec:EffectiveModels}, which also considers the general case in which molecules are not pinned.  

We will label the two resonant internal states of the molecule as $|\bar{0}\rangle$ and $|\bar{1}\rangle$\footnote{The bars indicate that these are microwave-dressed levels, as discussed in Sec.~\ref{sec:Microwave}.}, and assume that all other states are far-detuned on the scale of interactions and so may be neglected.  Further, let us assume that the states $|\bar{0}\rangle$ and $|\bar{1}\rangle$ have well-defined internal angular momentum projections $M=0$ and 1, respectively, for simplicity, though we stress that the assumption of well-defined angular momentum is not essential.  Two molecules, call them molecules $i$ and $j$, interact through the dipole-dipole interaction
\begin{align}
\label{eq:HDDI1}\hat{H}_{\mathrm{DDI}}&=\frac{\hat{\mathbf{d}}^{\left(i\right)}\cdot\hat{\mathbf{d}}^{\left(j\right)}-\left(e_{ij}\cdot\hat{\mathbf{d}}^{\left(i\right)}\right)\left(e_{ij}\cdot\hat{\mathbf{d}}^{\left(j\right)}\right)}{R_{ij}^3}\, ,
\end{align}
where $\mathbf{R}_{ij}$ is the vector connecting two molecules, $e_{ij}$ is a unit vector along $\mathbf{R}_{ij}$, and $\hat{\mathbf{d}}^{\left(i\right)}$ is the molecular dipole operator for molecule $i$.  For our purposes, it is useful to recast the dipole-dipole interaction as
\begin{align}
\label{eq:tenscont}\hat{H}_{\mathrm{DDI}}&=-\frac{\sqrt{6}}{R_{ij}^3} C^{\left(2\right)}\left(\mathbf{R}_{ij}\right)\cdot \left[\hat{\mathbf{d}}^{\left(i\right)}\otimes \hat{\mathbf{d}}^{\left(j\right)}\right]^{\left(2\right)}\, ,
\end{align}
where $C^{\left(2\right)}_m\left(\mathbf{R}_{ij}\right)=\sqrt{\frac{4\pi}{5}}Y^{\left(2\right)}_m\left(\mathbf{R}_{ij}\right)$ is an unnormalized spherical harmonic, $\left[\hat{\mathbf{d}}^{\left(i\right)}\otimes \hat{\mathbf{d}}^{\left(j\right)}\right]^{\left(2\right)}_q=\sum_{m=-1}^{1}\langle 1,m,1,{q-m}|2,q\rangle \hat{d}_m\hat{d}_{q-m}$ is a component of the rank-two decomposition of the tensor product of dipole operators with $\langle j_1,m_1, j_2,m_2|J,M\rangle$ a Clebsch-Gordan coefficient and $\hat{d}_{\pm 1}=\mp(\hat{d}_X\pm i \hat{d}_Y)/\sqrt{2}$, $\hat{d}_0=\hat{d}_Z$ spherical components of the dipole operator, and the dot denotes tensor contraction.  Expanding out the tensor contraction Eq.~\eqref{eq:tenscont}, we have
\begin{align}
\label{eq:compleDDI}\hat{H}_{\mathrm{DDI}}&=\hat{H}_{q=0}+\hat{H}_{q=\pm 1}+\hat{H}_{q=\pm2}\, ,
\end{align}
where
\begin{align}
\label{eq:q0}\hat{H}_{q=0}&=\frac{1-3\cos^2\theta_{ij}}{{R_{ij}}^3}\\
\nonumber &\times \left[\frac{\hat{d}^{\left(i\right)}_1\hat{d}^{\left(j\right)}_{-1}+\hat{d}^{\left(i\right)}_{-1}\hat{d}^{\left(j\right)}_1}{2}+\hat{d}^{\left(i\right)}_0\hat{d}^{\left(j\right)}_0\right]\, ,\\
\label{eq:q1}\hat{H}_{q=\pm 1}&=\frac{3}{\sqrt{2}}\frac{\sin\theta_{ij}\cos\theta_{ij}}{{R_{ij}}^3}\\
\nonumber &\times \left[\left(\hat{d}^{\left(i\right)}_1\hat{d}^{\left(j\right)}_0+\hat{d}^{\left(i\right)}_0\hat{d}^{\left(j\right)}_1\right)e^{-i\phi_{ij}}+\mathrm{h.c.}\right]\, ,\\
\label{eq:q2}\hat{H}_{q=\pm 2}&=-\frac{3}{2}\frac{\sin^2\theta_{ij}}{{R_{ij}}^3}\left[\hat{d}^{\left(i\right)}_1\hat{d}^{\left(j\right)}_1e^{-2i\phi_{ij}}+\mathrm{h.c.}\right]\,.
\end{align}
Here, $\theta_{ij}$ is the polar angle between $\mathbf{R}_{ij}$ and the quantization axis and $\phi_{ij}$ is the azimuthal angle in the $XY$ plane.  

For the 2D geometry of Fig.~\ref{fig:Levels}(b), $\theta_{ij}=\pi/2$, and so the interactions Eq.~\eqref{eq:q1} vanish identically.  The term Eq.~\eqref{eq:q0} conserves the internal and orbital angular momenta separately.  Hence, when projected into our basis $\left\{|\bar{0}\rangle,|\bar{1}\rangle\right\}$ of states with well-defined internal angular momentum, the most generic spin-spin coupling that can result is
\begin{align}
\label{eq:SpinEx}\hat{H}_{q=0}&=\frac{1-3\cos^2\theta_{ij}}{R^3}\\
\nonumber &\times \left[\frac{J_{\perp}}{2}\left(\hat{S}_{+}^{\left(i\right)}\hat{S}_{-}^{\left(j\right)}+\mathrm{h.c.}\right)+J_z\hat{S}_z^{\left(i\right)}\hat{S}_z^{\left(j\right)}\right]\, ,\\
&=\frac{1-3\cos^2\theta_{ij}}{R^3}\\
\nonumber &\times \left[J_{\perp}\left(\hat{S}_{x}^{\left(i\right)}\hat{S}_{x}^{\left(j\right)}+\hat{S}_{y}^{\left(i\right)}\hat{S}_{y}^{\left(j\right)}\right)+J_z\hat{S}_z^{\left(i\right)}\hat{S}_z^{\left(j\right)}\right]\, ,
\end{align}
where $\hat{S}_{\pm}, \hat{S}_x, \hat{S}_y,$ and $\hat{S}_z$ are pseudospin-1/2 operators in the effective spin space $\{|\uparrow\rangle, |\downarrow\rangle\}=\{|\bar{0}\rangle, |\bar{1}\rangle\}$ and we have ignored constant terms and single-spin terms~\cite{WHR_Review}.  The coupling constants $J_{\perp}$ and $J_z$ are set by dipole matrix elements, and in general are affected by external confinement.  The terms proportional to $J_{\perp}$ are responsible for the ``spin exchange" or ``state swapping"~\cite{PhysRevA.84.061605} dipolar interactions which were observed recently in the KRb experiment at JILA~\cite{Yan_Moses}, and the $J_z$ terms account for the fact that interactions between molecules in the $|\bar{0}\rangle$ state may be different from interactions between molecules in the $|\bar{1}\rangle$ state. 

In contrast to the $q=0$ term, Eq.~\eqref{eq:q2} does not conserve internal and external angular momentum separately, but transfers two units of angular momentum from the molecular rotation to the orbital motion or vice versa.  Projected into our two-state basis, these read
\begin{align}
\label{eq:prjqpm2}\hat{H}_{q=\pm 2}&=-\frac{3}{2}\frac{\sin^2\theta_{ij}}{R_{ij}^3}\left[J_{\Delta}\hat{S}_{+}^{\left(i\right)}\hat{S}_{+}^{\left(j\right)}e^{-2 i\phi_{ij}}+\mathrm{h.c.}\right]\, ,\\
\nonumber &=-3\frac{\sin^2\theta_{ij}}{R_{ij}^3}J_{\Delta}\Big[\cos\left(2\phi_{ij}\right)\left(\hat{S}_x^{\left(i\right)}\hat{S}_x^{\left(j\right)}-\hat{S}_y^{\left(i\right)}\hat{S}_y^{\left(j\right)}\right)\\
&+\sin\left(2\phi_{ij}\right)\left(\hat{S}_x^{\left(i\right)}\hat{S}_y^{\left(j\right)}+\hat{S}_y^{\left(i\right)}\hat{S}_x^{\left(j\right)}\right)\Big]\, .
\end{align}
Hence, the complete dipole-dipole interaction, Eq.~\eqref{eq:compleDDI}, projected into these two states allows for vast control over the X, Y, and Z components of the Heisenberg spin couplings via geometry and dipole matrix elements.  Models with unequal X and Y coupling strengths do not conserve the total magnetization.  Quantum spin models which do not conserve magnetization are of interest because they can generate quantum phases with no counterpart in magnetization-conserving systems, and also for their connection to Majorana fermions and other topological phenomena, see Sec.~\ref{sec:EffectiveModels}.  We stress that the $q=\pm2$ components, Eq.~\eqref{eq:q2}, which are responsible for the terms which do not conserve magnetization, Eq.~\eqref{eq:prjqpm2}, only contribute near a resonance such as in Fig.~\ref{fig:Resonance}(b).  In the remainder of this work, we will show how to engineer such resonances for symmetric top molecules, and also how to tune the effective spin-spin couplings $J_{\perp}$, $J_z$, and $J_{\Delta}$ (see Eqs.~\eqref{eq:XYZ} and \eqref{eq:XYZdeep} for the final spin model results).  Also, we relax many of the simplifying assumptions made in this section, such as the restriction that the molecules are pinned and that the molecular states have well-defined internal angular momentum projection.

%The terms Eqs.~\eqref{eq:q01}-\eqref{eq:q02} represent dipole-dipole interaction which do not change the internal state of the molecules.  The term Eq.~\eqref{eq:SPE} is a process in which two molecules exchange their internal state through simultaneous, in-phase dipole transitions $|M\rangle\to|M'\rangle$ and $|M'\rangle\to|M\rangle$.  These terms are responsible for the dipolar spin exchange observed in the recent KRb experiments at JILA~\cite{Yan_Moses,Hazzard}.  The final term Eq.~\eqref{eq:q2t} represents a resonance as drawn in Fig.~\ref{fig:Resonance}(b), in which both molecules make the transition $|M\rangle\to|M'\rangle$.

\section{Symmetric top molecules and their interaction with external fields}
\label{sec:STM}

In this section, we review the basic properties and energy scales of symmetric top molecules (STMs) and their interactions with both static and dynamic external fields.  A key result of this section is that STMs display a linear Stark effect, which is to say the energy varies linearly with the applied electric field strength at moderate fields.  A linear Stark effect has the consequence that a large portion of the dipole moment of an STM can be accessed with very modest electric fields.  We also show that the linear Stark effect can be used together with microwave dressing of low-lying rotational states to engineer level crossings in the single-molecule energy spectrum.  Such level crossings enable the realization of the resonances shown in Fig.~\ref{fig:Resonance}(b) that are key for the unconventional magnetism described in this work.

%\cs{The section needs a short summary paragraph describing what the different subsections are about}

\subsection{Rotational structure and interaction with static electric fields}
Polyatomic symmetric top molecules (STMs), of which methyl fluoride, {\MF}, is a canonical example, are defined by a doubly degenerate eigenvalue of the inertia tensor.  Such a doubly degenerate eigenvalue corresponds to a cylindrical symmetry of the molecule, see Fig.~\ref{fig:Levels}(a), and has key consequences for the rotational level structure of STMs.  The rotational degrees of freedom of a STM in the lowest electronic and vibrational state may be characterized by the rigid-rotor basis $\langle \omega_m|J,K,M\rangle=\sqrt{\frac{2J+1}{8\pi^2}}\mathcal{D}^{J\ast}_{MK}(\omega_m)$, where $J$ is the rotational quantum number, $M$ is the projection of rotation $\mathbf{J}$ on a space-fixed quantization axis, $K$ is the projection of $\mathbf{J}$ on the symmetry axis of the molecule, and $\mathcal{D}^{J}_{MK}(\omega_m)$ are the matrix elements of the Wigner $D$-matrix rotating the space-fixed frame to the molecule-fixed frame by the Euler angles $\omega_m$~{\cite{Zare}}, see Fig.~\ref{fig:Levels}(a).  The corresponding rotational eigenenergies are $E_{JKM}=B_0J(J+1)+(A_0-B_0)K^2$, where the rotational constants $B_0\approx25$GHz, $A_0\approx 155$GHz for {\MF}.  Diatomic $^1\Sigma$ molecules, such as the alkali dimers, cannot have a projection of $\mathbf{J}$ on the body axis, and so $K=0$ identically.

Just as the isotropy of space requires that the states with differing projections $M$ of $\mathbf{J}$ onto a space-fixed axis are degenerate in the absence of external fields, the cylindrical symmetry of STMs requires that states with opposite projection $\pm K$ of $\mathbf{J}$ onto a molecule-fixed axis are degenerate.  Corrections to the rigid rotor approximation in the vibration-rotation Hamiltonian, such as the well-known inversion of ammonia, can cause mixing of the $K$ levels and result in a splitting of this degeneracy.  For simplicity of discussion we will focus on molecules such as {\MF} which do not have an inversion splitting in the body of this paper, though we will revisit this issue at the end of the next subsection.

The presence of a nonzero molecule-frame projection of rotational angular momentum $K$ in STMs means that STMs can display a {linear response} to an externally applied static electric field.  This is in stark contrast to the quadratic {response} exhibited by $\Sigma$-state molecules such as the alkali metal dimers~\cite{Wall_Maeda_13}.  In particular, in a static electric field of strength $E_{\mathrm{DC}}\ll B_0/d$ defining the quantization axis, with $d$ the permanent dipole moment, the matrix elements of the dipole operator along space-fixed spherical direction $p$, $\hat{d}_p$, take the form of a spherical tensor with reduced matrix element $\langle J,K'||\hat{\mathbf{d}}||J,K\rangle=dK\sqrt{\frac{2J+1}{J(J+1)}}\delta_{K,K'}$~\cite{Wall_Maeda_13}.  Hence, STMs in this field regime display a linear Stark effect with eigenenergies {$E_{JKM}=-dKME_{\mathrm{DC}}/[J(J+1)]$}.

The strong coupling of STMs to external fields enables them to be effectively decelerated by electric fields~\cite{PhysRevLett.112.013001}, and is the basis of opto-electrical cooling, a novel route to bring generic STMs to quantum degeneracy~\cite{PhysRevLett.107.263003,oec}.  Furthermore, the nonzero reduced matrix element of the dipole operator within a rotational state manifold enables STMs to simulate the physics of magnetic dipoles and quantum magnetism with greatly enhanced dipolar interaction energies~\cite{Wall_Maeda_13}.  In Ref.~\cite{Wall_Maeda_13}, we showed how this correspondence between STMs with rotational quantum number $J$ and an elemental quantum magnet with spin $J$ gives rise to long-range and anisotropic spin models.  In what follows, we introduce microwave dressing of rotational states as an additional handle with which to modify the forms and relative strengths of interactions that appear in such effective spin models.

\subsection{Microwave dressing of symmetric top molecules}
\label{sec:Microwave}

\begin{figure}[tbp]
\centerline{\includegraphics[width=\columnwidth]{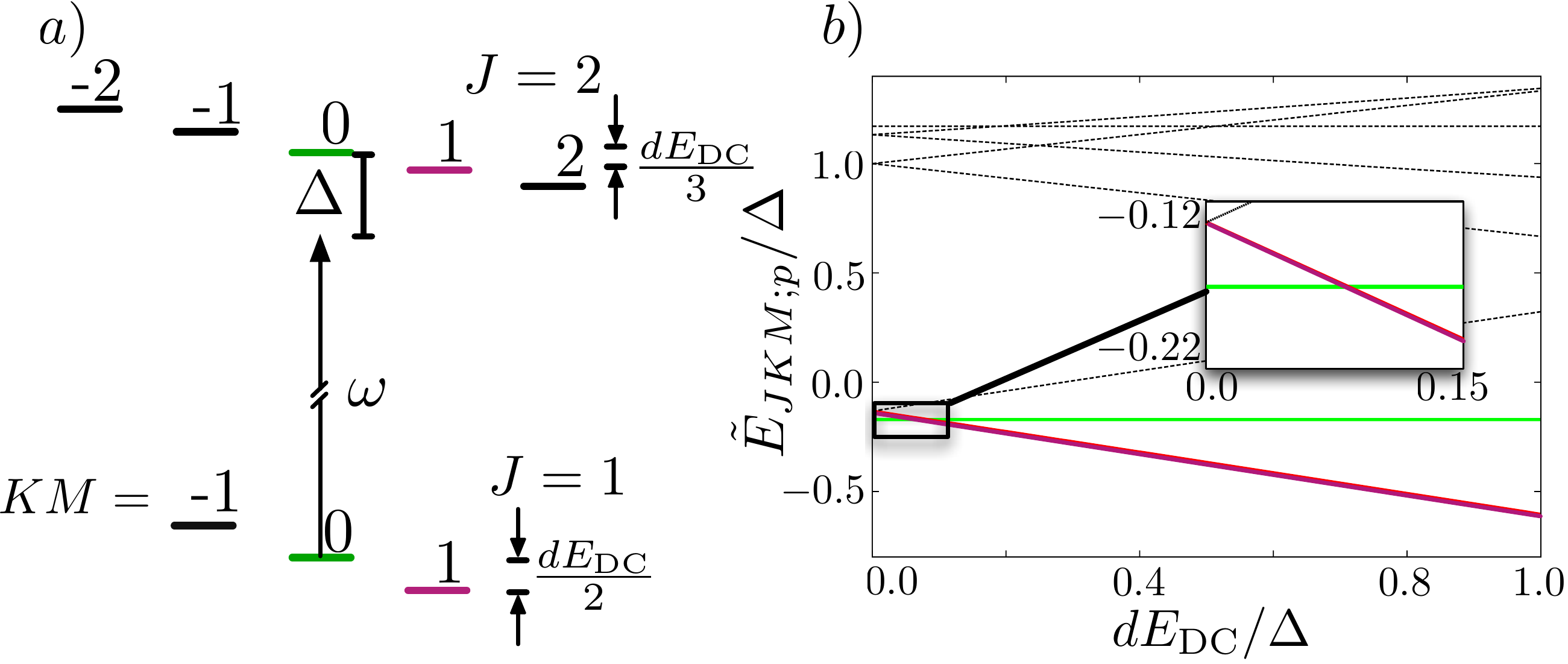}}
\caption{\label{fig:p=0} 
\emph{Microwave-dressing of symmetric top molecules with linear polarization} (a) The manifolds $(J,K)=(1,1)$ and $(2,1)$ have different linear Stark effects.  Coupling by a $z$-polarized microwave of frequency $\omega$ and detuning $\Delta$ generates dressed state $|\bar{0}\rangle$ ($|\bar{1}\rangle$) in green (purple).  (b) Combined DC and AC Stark shift cause a level crossing of $|\bar{0}\rangle$ and $|\bar{1}\rangle$ ($\Omega/\Delta=1$).  Inset: enlargement of crossing.}
\end{figure}

Microwave radiation couples together neighboring rotational states of a molecule when the frequency of the radiation is near-resonant with the rotational energy level difference.  For simplicity, we first consider applying a microwave field $\mathbf{E}_{\mathrm{AC}}$ with linear polarization along the space-fixed quantization axis, $\boldsymbol{\varepsilon}_{\mathrm{AC}}=\mathbf{e}_Z$, see Fig.~\ref{fig:Levels}(b), which is red-detuned an amount $\Delta\ll B_0$~\footnote{We adopt units in which $\hbar=1$ unless specified otherwise.} from resonance with the $|J,K,0\rangle\to|J+1,K,0\rangle$ transition, as shown in Fig.~\ref{fig:p=0}(a).  While the frequency of this transition in {\MF} is larger than the corresponding rotational transition in the alkali dimers, the wavenumber of the transition $k\approx 2(J+1)B_0/(h c)$ is much less than $1/a$, with $a$ the average separation between molecules, of order a few hundred nanometers for typical optical lattices.  Hence, we neglect the spatial dependence of the microwave field.  Applying the rotating wave approximation and transforming to the Floquet picture~\cite{Shirley}, the quasienergies are obtained by solving the Schr\"odinger equation for fixed $M$ with the $2\times 2$ Hamiltonians~\footnote{States in the $(J+1,K)$-manifold which do not couple to any states in the $\left(J,K\right)$-manifold have trivial $1\times 1$ Hamiltonians and we ignore them.}:
\begin{align}
\label{eq:HJKM}\hat{H}_{JKM}&=\left(\begin{array}{cc} -\frac{dKME_{\mathrm{DC}}}{J\left(J+1\right)}&-\Omega_{JKM}\\ -\Omega_{JKM}&\Delta-\frac{dKME_{\mathrm{DC}}}{\left(J+1\right)\left(J+2\right)}\end{array}\right)\, ,
\end{align}
where 
\begin{align}
\Omega_{JKM}\equiv \Omega\left\{\frac{[(J+1)^2-K^2][(J+1)^2-M^2]}{(J+1)^2(2J+1)(2J+3)}\right\}^{1/2}\, ,
\end{align}
with the Rabi frequency $\Omega\equiv dE_{\mathrm{AC}}$.  Single-particle eigenstates of Eq.~\eqref{eq:HJKM} in the rotating frame will be denoted by an overbar, e.g., $|\bar{0}\rangle$.

In the perturbative regime where $\Omega,dE_{\mathrm{DC}}\ll \Delta$, the quasienergies are split into manifolds $\tilde{E}_{JKM;\pm}$ separated by roughly $\Delta$, see Fig.~\ref{fig:p=0}(b).  The $M$ dependence of the off-diagonal components $\Omega_{JKM}$ introduces an effective tensor shift between states of different $M$ which is proportional to $\Omega^2$, similar to the microwave-induced quadratic Zeeman effect in spinor Bose gases~\cite{PhysRevA.73.041602,PhysRevA.75.053606}.  Including the static field $E_{\mathrm{DC}}$ can cause two such quasienergy levels with different $M$ to cross as the static field energy $dE_{\mathrm{DC}}$ becomes of the order of the effective tensor shift, as shown for the case of the $(J,K)=(1,1)\to(2,1)$ transition in Fig.~\ref{fig:p=0}(a).  The ability to engineer \emph{generic} quasienergy level crossings by tuning the static electric field strength is a consequence of the linear Stark effect exhibited by STMs.  The Stark effect in $^1\Sigma$ molecules is quadratic, and so shifts all levels with identical $J$ and $|M|$ in the same fashion.

%\cs{This might be be a bit unclear to the non-molecular-physicist reader.  Can you please clarify?  You want to say $1\Sigma$ molecules are better/worse for certain purposes than STMs.  So try to bring that point out here.}For $^1\Sigma$ molecules, an external electric field shifts all levels with identical $J$ and $|M|$ the same, and so a moderate electric field has the same effect on the internal levels as the microwave tensor shift.  

\begin{figure}[tbp]
\centerline{\includegraphics[width=\columnwidth]{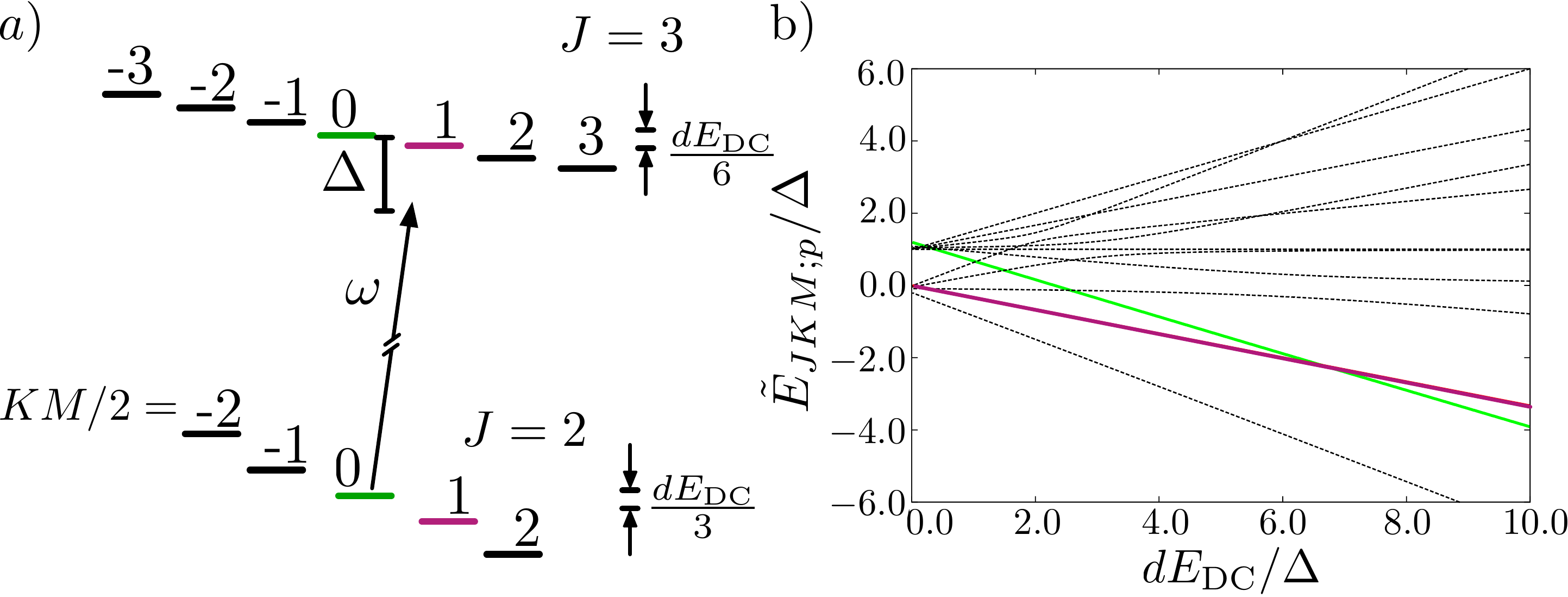}}
\caption{\label{fig:p=1} 
\emph{Microwave-dressing of symmetric top molecules with circular polarization} (a) The manifolds $(J,K)=(2,2)$ and $(3,2)$ display different linear Stark effects.  Coupling by a right-circularly polarized microwave of frequency $\omega$ and detuning $\Delta$ generates dressed state $|\bar{0}\rangle$ ($|\bar{1}\rangle$) in green (purple).  (b) A level crossing of $|\bar{0}\rangle$ and $|\bar{1}\rangle$ occurs when the differential DC Stark shift becomes on the order of the detuning ($\Omega/\Delta=1$).}
\end{figure}

Level crossings can also be engineered outside of the perturbative regime, as well as for arbitrary polarization and rotational quantum number $J$.  As an example, we consider the transition $(J,K)=(2,2)\to(3,2)$ with right-circularly polarized light in Fig.~\ref{fig:p=1}.   Panel (b) of Fig.~\ref{fig:p=1} shows two levels which cross outside of the perturbative regime.  Here, the linear Stark energy must overcome not only the effective tensor shift, but also the detuning $\Delta$ of the microwave field from resonance.  In what follows, we will denote the parametric relationship of the Rabi frequency $\Omega$ and the electric field at a quasienergy level crossing as $\tilde{\Omega}\left(E_{\mathrm{DC}}\right)$.  

 In our analysis of the field dressing of STMs we have neglected hyperfine structure.  Though the hyperfine structure of STMs is complicated~\cite{Wall_Maeda_13}, a single hyperfine component may be selected via a strong magnetic field, similarly to the alkali dimers~\cite{Ospelkaus_Ni_10}.  Alternatively, working at microwave detuning large compared to the typical hyperfine splittings, $\Delta\gg E_{\mathrm{hfs}}\approx 10$kHz for {\MF}, one can address all hyperfine states equally with a readily achievable microwave power on the order of tens of W/cm$^2$.  While we have focused on polyatomic STMs in which all states with a given $J$ and $K$ are degenerate in zero DC field, we expect similar level crossings in other systems with a linear Stark effect but no zero-field degeneracy, such as the Lambda doublet of OH~\cite{OH}, its fermionic analog OD~\cite{OD}, or other species with non-zero projection of orbital angular momentum along the symmetry axis of the molecule $|\Lambda|>0$.  Generally, one can take the detuning $\Delta$ much larger than any fine energy scale which is not to be resolved and simply rescale the static field energy $dE_{\mathrm{DC}}$ and the Rabi frequency $\Omega$ accordingly.

%We stress that the ability to rescale both the static field energy $dE_{\mathrm{DC}}$ and the microwave field energy $dE_{\mathrm{AC}}$ by $\Delta$ means that $\Delta$ can be taken much larger than any energy scale smaller than the rotational structure, such as hyperfine structure or inversion/Lambda doublets.   Hence, our dressing scheme applies even in the presence of such structure.

\subsection{Dipole-dipole interactions in microwave-dressed states}
\label{sec:DDImicrowave}
We now turn to the effective dipole-dipole interactions (Eq.~\eqref{eq:compleDDI}) in the microwave-dressed states.  The components of the dressed states $|\bar{0}\rangle$ and $|\bar{1}\rangle$ in the $|J+1,K,M\rangle$ manifold oscillate in time with frequency $\omega$.  Hence, the dipole moments of the dressed states contain both static and time-oscillating pieces.  While the oscillating terms time-average to zero for a single molecule, the dipole-allowed exchange of rotational quanta for two molecules can be resonant due to the two dipoles oscillating in phase~\cite{PhysRevLett.103.155302,Yan_Moses}.  In more detail, let us consider the Floquet-picture eigenstates in the presence of a microwave with spherical polarization $p=0$, which couples states $|JKM\rangle$ to $|(J+1)KM\rangle$, see Fig.~\ref{fig:p=0}(a).  Consider two such eigenstates
\begin{align}
|M\rangle&=a|JKM\rangle+be^{-i\omega t}|\left(J+1\right)KM\rangle\, ,\\
|M'\rangle&=c|JKM'\rangle+de^{-i\omega t}|\left(J+1\right)KM'\rangle\, .
\end{align}
Because we are interested in levels which cross, we assume that $M\ne M'$.  From Eq.~\eqref{eq:HJKM}, two levels with the same $M$ do not cross except in trivial cases.  The matrix elements of the dipole operators are
\begin{align}
\label{eq:dipstart}\langle M|\hat{d}_p|M\rangle&=a^2D_{J,J;K}^{M,M;p}+b^2D_{J+1,J+1;K}^{M,M;p}\\
\nonumber &+ab\left(D_{J,J+1;K}^{M,M;p}e^{-i\omega t}+D_{J+1,J;K}^{M,M;p}e^{i\omega t}\right)\, ,\\
\langle M'|\hat{d}_p|M'\rangle&=c^2D_{J,J;K}^{M',M';p}+d^2D_{J+1,J+1;K}^{M',M';p}\\
\nonumber &+cd\left(D_{J,J+1;K}^{M',M';p}e^{-i\omega t}+D_{J+1,J;K}^{M',M';p}e^{i\omega t}\right)\, ,\\
\label{eq:dipend}\langle M|\hat{d}_p|M'\rangle&=acD_{J,J;K}^{M,M';p}+bdD_{J+1,J+1;K}^{M,M';p}\\
\nonumber &+ade^{-i\omega t}D_{J,J+1;K}^{M,M';p}+bce^{i\omega t}D_{J+1,J;K}^{M,M';p}\, ,
\end{align}
where
\begin{align}
D_{J',J;K}^{M',M;p}&\equiv \left(-1\right)^{M'-K}\sqrt{\left(2J'+1\right)\left(2J+1\right)}\\
\nonumber &\times \tj{J'}{-M'}{1}{p}{J}{M}\tj{J'}{-K}{1}{0}{J}{K}\, ,
\end{align}
are the dipole matrix elements in pure rotational STM states.  Here, we recall that $J$ is the rotational principal quantum number, $M$ is the projection of rotation on a space-fixed quantization axis, and $K$ is the projection of the rotational angular momentum on the symmetry axis of the molecule.  %\cs{Go back over indices, maybe also adding clarification}

In order to find the effective dipole-dipole interactions, we take the matrix elements of the dipole operators in Eq.~\eqref{eq:q0}-\eqref{eq:q2} using the matrix elements of Eq.~\eqref{eq:dipstart}-\eqref{eq:dipend} and then perform the long-time average.  Here, ``long" time refers to a time which is long compared to the period of the microwave field.  The long-time average is justified by the fact that the characteristic timescales of the translational motion of molecules are orders of magnitude longer than the period of the dressing field. The resulting time-averaged interactions for our two example polarization schemes are discussed in Appendix~\ref{appendix:A}, and specific numerical examples of interactions for these two polarizations are given in the next section.  The only assumption we use in this work is that the dipole moments of two states near a level crossing only have static components along a single space-fixed spherical direction.  Practically, the microwave field can contain either $p=\pm 1$ components or $p=0$ components, but not both.  This is equivalent to the statement that terms which transfer only a single molecule between dressed state components are all proportional to $\sin\theta\cos\theta$, and so vanish in the geometry of Fig.~\ref{fig:Levels}.  The requirement of only a single microwave frequency  is in contrast to proposals with $^1\Sigma$ molecules, which often require precise frequency and polarization control of {multiple microwaves}~\cite{Gorshkov_Manmana_11,Gorshkov_Manmana_11b,PhysRevB.87.081106,doi:10.1080/00268976.2013.800604}.

%and are the interactions which appear in our effective many-body models derived in the next section.

\section{Unconventional Hubbard and Spin models with symmetric top molecules}
\label{sec:EffectiveModels}

%\cs{The section needs a short summary paragraph describing what the different subsections are about.  Actually this is the most exciting section in the paper.  So we want the excitement to come across.  The section/subsection titles could also be sexier}

In this section we incorporate the single-particle physics discussed in the previous section into an interacting many-body description in second quantization.  Following a translation of the many-body problem to a Hubbard-type lattice model for the lowest lattice band, we then show how limiting cases of this description, for example when the molecules are pinned to lattice sites, leads to spin models with unconventional magnetic couplings.  Our main results are Eq.~\eqref{eq:MBModel}, the most complete Hubbard-type description of the physics of STMs near a quasienergy level crossing, and Eqs.~\eqref{eq:XYZdeep} and Eqs.~\eqref{eq:KitaevXY}, which are the reductions to the Heisenberg XYZ and XY models, respectively. 

\subsection{Second-quantized description of physics near a quasienergy level crossing}
\label{sec:lattmodel}
%\cs{Can we come up with a better name than "lattice model?"}

In order to derive an effective model for the microwave-dressed STMs trapped in an optical lattice, we use the standard prescription~\cite{Jaksch_PRL_1998} of expanding the field operator in the second quantized representation of the Hamiltonian in a basis of Wannier functions and keep only the terms corresponding to the lowest band of the lattice.  Additionally, in what follows we will assume hard-core particles, which can be either bosons or fermions.  By hard-core we specifically mean no more than one molecule can simultaneously occupy a given lattice site irrespective of internal state considerations.  Such a constraint can arise either from a large positive elastic interaction energy, or from rapid inelastic losses via the quantum Zeno effect.  The quantum Zeno effect has been shown to enforce a hard-core constraint for KRb, where {two-body losses} are due to chemical reactions, and gives rise to lifetimes which are long compared to the typical time scales of interactions~\cite{Yan_Moses,Zeno}.  Because our {scheme populates multiple dressed states consisting of different rotational levels}, molecules {undergo possibly rapid rotationally inelastic processes at short range} which will cause a loss of molecules from the trap even if the molecules themselves are chemically stable.  The numerical examples given in this work have sufficiently large elastic on-site interactions that we do not need to worry about the nature of short-range inelastic collisions, and we can attribute the hard-core constraint to elastic interactions alone.

\begin{figure}[tbp]
\centerline{\includegraphics[width=1.0\columnwidth]{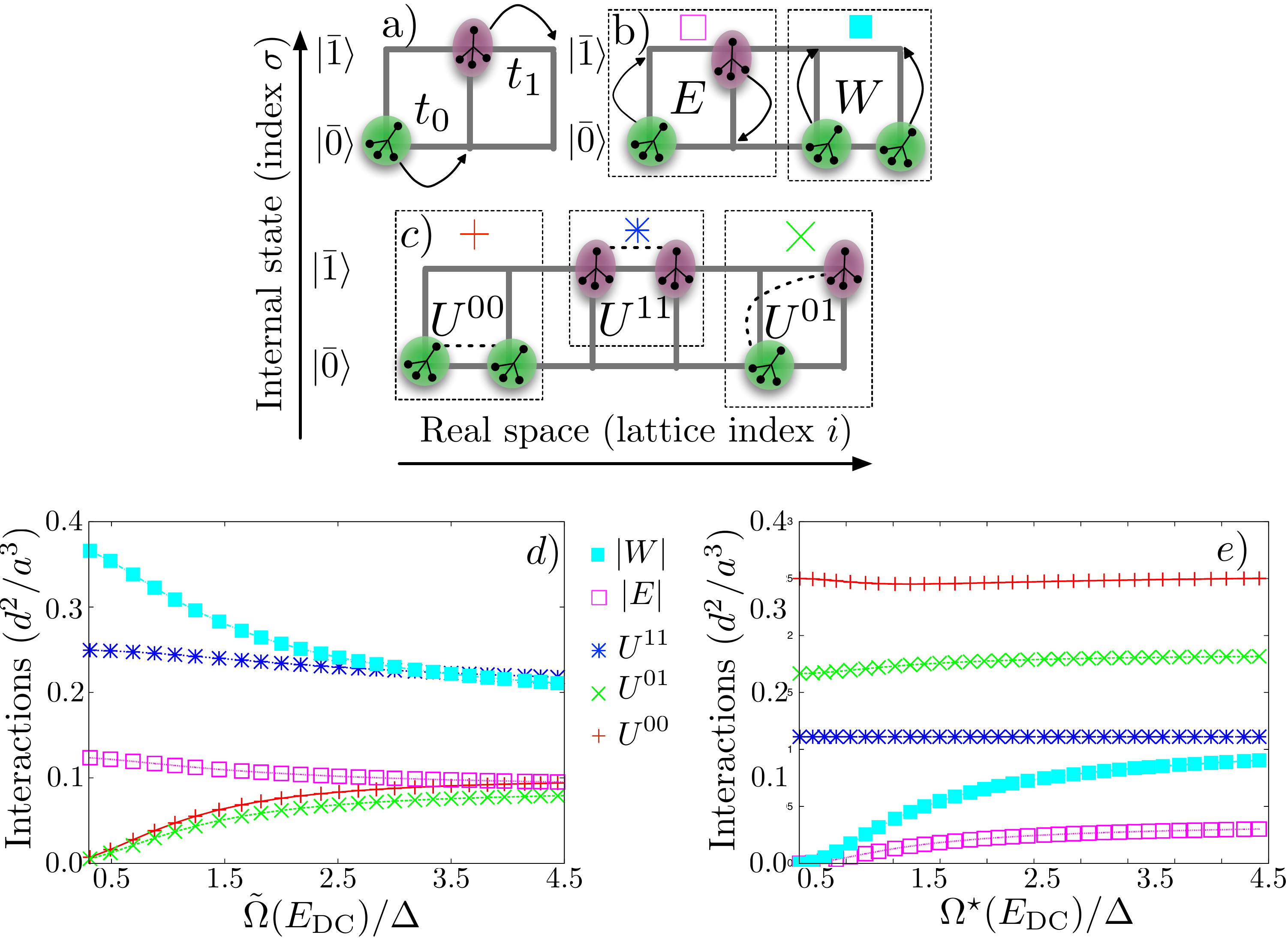}}
\caption{\label{fig:ID}  \emph{Interaction processes in the effective lattice Hamiltonian.} The two internal states $|\bar{0}\rangle$ and $|\bar{1}\rangle$ may be viewed as an discrete spatial degree of freedom, e.g.~a ladder. (a) Tunneling rates $t_{\sigma}$ depend on the internal state due to polarizability anisotropy~\cite{Wall_Maeda_13,neyenhuis:anisotropic_2012}.  (b) $E$ and $W$ interactions change the internal state of the molecules.  $E$ processes preserve the number in each internal state, $W$ processes change it by $\pm 2$.  (c) $U$ interactions preserve the internal state of the molecules.   (d) Nearest-neighbor Hubbard parameters ($W\equiv W_{i,i+1}$ etc.) of the many-body model Eq.~\eqref{eq:MBModel} with ($E_{\mathrm{DC}}$-dependent) Rabi frequency $\tilde{\Omega}$ at the level crossing in Fig.~\ref{fig:p=0}.  (e) Same as (d) for the level crossing in Fig.~\ref{fig:p=1}.  Symbols correlate processes to Panels (a)-(c).
}
\end{figure}

For two dressed states $\sigma\in\left\{\bar{0},\bar{1}\right\}$ which are separated from all others by an energy large compared to the characteristic dipole-dipole energy scale, an expansion of the full many-body description in terms of the lowest band Wannier functions reads
\begin{align} \label{eq:MBModel}
\nonumber \hat{H}&=-\sum_{\langle i,j\rangle,\sigma}t_{\sigma}\hat{a}_{i\sigma}^{\dagger}\hat{a}_{j\sigma}+\delta \sum_{i}\hat{n}_{i\bar{1}}\\
\nonumber &+\frac{1}{2}\sum_{i,j,i\ne j}[E_{i,j}\hat{S}^+_{i}\hat{S}^{-}_j+W_{i,j}\hat{S}^+_i\hat{S}^+_j+\mathrm{h.c.}]\\
&+\frac{1}{2}\sum_{\sigma,\sigma', i,j,i\ne j}U^{\sigma\sigma'}_{i,j}\hat{n}_{i\sigma}\hat{n}_{j\sigma}\, .
\end{align}
Here, $\hat{a}_{i\sigma}$ destroys a STM in Wannier state $w_{i\sigma}(\mathbf{r})$ centered at site $i$, $\hat{n}_{i\sigma}=\hat{a}_{i\sigma}^{\dagger}\hat{a}_{i\sigma}$, and $\hat{S}^+_i=\hat{a}_{i\bar{0}}^{\dagger}\hat{a}_{i\bar{1}}$, $\hat{S}^-_i=(\hat{S}_i^+)^{\dagger}$ are spin-1/2 operators.  In order, the terms in Eq.~\eqref{eq:MBModel} are state-dependent tunneling $t_{\sigma}$ of molecules between neighboring lattice sites $\langle i,j\rangle$\footnote{As has been discussed in previous work considering long-range interacting systems~\cite{PhysRevA.88.023605}, truncation of tunneling to the nearest neighbor distance may not represent a consistent order of approximation when the interactions have infinite range.  Instead, one can keep the tunneling between sites to a range $r_t$ and the interactions out to a range $r_U$ such that all discarded terms in the Hamiltonian have a bounded magnitude.}; a single-particle energy offset $\delta$ of state {$|\bar{1}\rangle$} with respect to state $|\bar{0}\rangle$; state-exchanging collisions $E_{i,j}$ of molecules at sites $i$ and $j$; state-transferring collisions $W_{i,j}$ which transform two molecules in state $|\bar{0}\rangle$ at sites $i$ and $j$ into the state $|\bar{1}\rangle$ and vice versa;  and state-preserving collisions $U_{i,j}^{\sigma\sigma'}$ between molecules in states $\sigma$ and $\sigma'$ at lattice sites $i$ and $j$, respectively.  A schematic view of the processes in Eq.~\eqref{eq:MBModel} is given in Fig.~\ref{fig:ID}(a)-(c).  Note that Eq.~\eqref{eq:MBModel} applies to any 2D lattice geometry.

%\cs{I would at least have a footnote that longer range hopping may be necessary, as well as other terms, upon consistent truncation to a time/energy scale, as we've discussed in the past.  I think you should mention molecular Hubbard models here and explain how this one differs (by the W term only).  Then explain what the W term is for.}

%\cs{Generally I feel this subsection needs a bit of work.  Assume a person has never seen a molecular Hubbard Hamiltonian and possibly no Hubbard Hamiltonian at all.  Connect back to our earlier work on STMs as well as $1\Sigma$ models (field regimes paper).}  

%\cs{I think you can move this paragraph to the first of this section, and slightly adjust the language so that it leads into the Hamiltonian.  Placed where it is now it breaks the flow of the discussion.}  I

%\cs{This should be part of the paragraph after Eq. (18)} 

%\cs{Compare these terms to magnetism -- direct and exchange terms.  Explain extra term W.  Some readers will have a magnetism perspective.}

The Hamiltonian Eq.~\eqref{eq:MBModel} bears a strong resemblance to the molecular Hubbard Hamiltonian (MHH) which has been derived for $^1\Sigma$ molecules in optical lattices~\cite{Wall_Carr_NJP,Wall_PRA_2010,PhysRevA.88.023605}, and many of the terms here have the same meaning as in the MHH.  In particular, the interaction terms $U_{i,j}^{\sigma\sigma'}$ correspond to the direct terms $\langle \sigma \sigma'|\hat{H}_{\mathrm{DDI}}|\sigma\sigma'\rangle$, and the interaction terms $E_{i,j}$ correspond to the exchange terms $\langle \sigma \sigma'|\hat{H}_{\mathrm{DDI}}|\sigma'\sigma\rangle$, where $\hat{H}_{\mathrm{DDI}}$ is given in Eq.~\eqref{eq:HDDI1}.  Expressed in the language of quantum magnetism for a spin-1/2 encoded in the states $|\left\{|\bar{0}\rangle, |\bar{1}\rangle\right\}$, the $U$ terms correspond to $\hat{S}^z\hat{S}^z$ or Ising-type interactions and the $E$ terms correspond to $\hat{S}^+\hat{S}^-$ or spin-exchange-type interactions~\cite{Gorshkov_Manmana_11}.  The new terms here, which have no counterpart in the MHH description of $^1\Sigma$ molecules, are the $W_{i,j}$ terms, which correspond to $\langle \sigma \sigma|\hat{H}_{\mathrm{DDI}}|\sigma'\sigma'\rangle$.  These terms correspond to $\hat{S}^+\hat{S}^+$-type interactions in the spin language, and are absent from typical Heisenberg XXZ-type models of quantum magnetism, including those realized with $^1\Sigma$ polar molecules~\cite{Gorshkov_Manmana_11}.  

All of the interaction coefficients $U$, $E$, and $W$ may be tuned by adjusting the static and microwave field dressing strengths, ensuring that the two quasienergy levels remain near resonance.  The magnitudes of the Hubbard parameters for the specific level crossings in Figs.~\ref{fig:p=0} and \ref{fig:p=1} are displayed in Fig.~\ref{fig:ID}(d)-(e) as a function of the $E_{\mathrm{DC}}$-dependent Rabi frequency at the level crossing, $\tilde{\Omega}(E_{\mathrm{DC}})$, with analytic expressions given in the appendix. For these dressing schemes, the Hubbard parameters $U$ and $E$ are overlaps of the $q=0$ component of the dipole-dipole potential, Eq.~\eqref{eq:q0}, in the basis of Wannier functions~\cite{Wall_Carr_CE}, while the $W$ terms involve overlaps of the $q=\pm 2$ components of the dipole-dipole potential, Eq.~\eqref{eq:q2}.  All dipolar parameters $U$, $E$, and $W$ have an approximately $1/|i-j|^3$ decay between lattice sites, and the $W$ terms additionally feature a dependence on the azimuthal angle $\phi$.  Other dressing schemes, for example those involving both $p=\pm 1$ polarizations, divide this angular dependence between $U$, $E$, and $W$.

The Hamiltonian Eq.~\eqref{eq:MBModel} has a U(1) symmetry generated by the total number operator $\hat{N}=\hat{N}_0+\hat{N}_1$ with $\hat{N}_{\sigma}= \sum_{i}\hat{n}_{i\sigma}$.  The $W$ term breaks number conservation within each internal state, but preserves the parities defined by $\hat{P}_{\sigma}=\exp(-i\pi \hat{N}_{\sigma})$.  Due to the U(1) symmetry, the two parities are redundant, both being proportional to $\hat{P}=\exp[-\frac{i\pi}{2}(\hat{N}_0-\hat{N}_1)]$, which is the parity of the number difference between internal states.  Hence, the internal symmetry of the model Eq.~\eqref{eq:MBModel} is U(1)$\times \mathbb{Z}_2$~\cite{PhysRevB.84.094503}.  We can interpret the $W$ term as being a hopping of pairs between two quantum wires or layers, where the wire indices correspond to the dressed states of the molecule, see Fig.~\ref{fig:ID}.  Due to the fact that exchange of rotational quanta only occurs when the dipoles oscillate in phase and the particular geometry, dipolar excitation of a single molecule is forbidden.  Single excitation processes which break the $\mathbb{Z}_2$ symmetry can be included systematically by other choices of geometry or field polarization, see Sec.~\ref{sec:DDImicrowave} and the appendix.  

\subsection{Mapping to a pure spin model}
In ultracold gases it is often easier to achieve low temperatures for the internal degrees of freedom even when the motional degrees of freedom remain hot.  As an example, a collection of molecules all prepared in the same quantum state has zero effective spin temperature.  Provided that the motional temperature of the molecules is lower than the rotational excitation energy (typically on the order of a few hundred milliKelvin), the spin and motional temperatures are effectively decoupled, and only the former is important for the dynamics.  Hence, a natural first step for many-body physics is to {freeze} the motional degrees of freedom by loading into a deep optical lattice and consider the dynamics of only the internal degrees of freedom~\cite{Yan_Moses}.  In the limit in which the quasi-2D confinement is so deep that the tunneling is negligible, Eq.~\eqref{eq:MBModel} becomes a long-range and anisotropic spin model
\begin{align}
\nonumber \hat{H}&=\textstyle\sum_{i,j,i\ne j}\Big[\left(E_{i,j}+W^{\mathcal{R}}_{i,j}\right)\hat{S}^x_i\hat{S}^x_j+\left(E_{i,j}-W^{\mathcal{R}}_{i,j}\right)\hat{S}^y_i\hat{S}^y_j\\
\nonumber&\textstyle-W^{\mathcal{I}}_{i,j}(\hat{S}^x_i\hat{S}^y_j+\hat{S}^y_i\hat{S}^x_j)+\left(U^{00}_{i,j}+U^{11}_{i,j}-2U^{01}_{i,j}\right)\hat{S}^z_i\hat{S}^z_j\Big]\\
\label{eq:XYZ}&\textstyle+\sum_ih_i\hat{S}^z_i\, ,
\end{align}
where $h_i=\delta+\frac{1}{4}\sum_{j,j\ne i}\left(U^{00}_{i,j}-U^{11}_{i,j}\right)\sum_{\sigma}\hat{n}_{j\sigma}$ is the effective magnetic field at site $i$, we have ignored a constant term, and $W^{\mathcal{R}}_{i,j}$ ($W^{\mathcal{I}}_{i,j}$) is the real (imaginary) part of $W_{i,j}$, the $\hat{S}^+_i\hat{S}^+_j$ coupling.  Again, we would like to stress that Eq.~\eqref{eq:XYZ} is defined on any 2D lattice geometry.  The Hamiltonian Eq.~\eqref{eq:XYZ} does not conserve magnetization {due to the non-zero $W^{\mathcal{R}}$ and $W^{\mathcal{I}}$ terms easily accessible in our scheme}, in contrast to the XXZ models realized with alkali dimer molecules~\cite{Gorshkov_Manmana_11,Yan_Moses}.  In deep optical lattices, where the Wannier functions become well-localized~\cite{Wall_Carr_CE}, the dipolar coupling constants can be approximated as
\begin{align}
\nonumber U^{\sigma\sigma'}_{i,j}&\approx \frac{U^{\sigma\sigma'} }{\left|i-j\right|^3}\, ,\;\;E_{i,j}&\approx \frac{E}{\left|i-j\right|^3}\, ,\;\; W_{i,j}&\approx W\frac{e^{-2i\phi_{i,j}}}{\left|i-j\right|^3}\, ,
\end{align}
where $\phi_{i,j}$ is the angle between the vector connecting sites $i$ and $j$ and the $x$ axis, and $U$, $E$, and $W$ are related to geometrical factors and expected dipole moments.  With these approximations, we can rewrite Eq.~\eqref{eq:XYZ} as
\begin{widetext}
\begin{align}
\nonumber \hat{H}&=\sum_{i,j,i\ne j}\frac{1}{\left|i-j\right|^3}\Big[\left(E+W\cos2\phi_{i,j}\right)\hat{S}^x_i\hat{S}^x_j +\left(E-W\cos2\phi_{i,j}\right)\hat{S}^y_i\hat{S}^y_j+W\sin 2\phi_{i,j}(\hat{S}^x_i\hat{S}^y_j+\hat{S}^y_i\hat{S}^x_j)\\
\label{eq:XYZdeep}&+\left(U^{00}+U^{11}-2U^{01}\right)\hat{S}^z_i\hat{S}^z_j\Big]+\sum_ih_i\hat{S}^z_i\, ,
\end{align}
\end{widetext}
which makes the spatial anisotropy of the model more explicit.

Some simplification of Eq.~\eqref{eq:XYZ} occurs in one spatial dimension (1D), which corresponds, e.g., to taking a single row of the 2D square lattice shown in Fig.~\ref{fig:Levels}(b).  Such a reduction can be performed in experiment by applying electric field gradients to select a single row of a 2D optical lattice.  In a 1D geometry, we can always choose coordinates such that the $x$ axis lies along the lattice direction, and so $W^{\mathcal{I}}_{i,j}$ vanishes and $W^{\mathcal{R}}_{i,j}$ is a monotonically decreasing function of $|i-j|$.  Here, Eq.~\eqref{eq:XYZ} reduces to a spin-1/2 XYZ Heisenberg model in a longitudinal field
\begin{align}
\hat{H}_{XYZ}&=\sum_{i,j,i\ne j}\left[J^X_{i,j}\hat{S}^x_i\hat{S}^x_j+J^Y_{i,j}\hat{S}^y_i\hat{S}^y_j+J^Z_{i,j}\hat{S}^z_i\hat{S}^z_j\right]\\
\nonumber &+\sum_i h_i\hat{S}^z_i\, ,
\end{align}
where $J^X_{i,j}=(E+W)/|i-j|^3$, $J^Y_{i,j}=(E-W)/|i-j|^3$, and $J^Z_{i,j}=(U^{00}+U^{11}-2U^{01})/|i-j|^3$.   Hence, the degree of spin anisotropy is tunable by changing the ratio between $E$ and $W$, see Fig.~\ref{fig:ID}.  The phase diagram of the nearest-neighbor version of this model has been investigated recently in Ref.~\cite{PhysRevLett.111.205302}, displaying Berezinsky-Kosterlitz-Thouless, Ising, first-order, and commensurate-incommensurate phase transitions.  Further, considering that the coefficient of $\hat{S}^z_i\hat{S}^z_j$ vanishes~\footnote{The coefficient of $\hat{S}^z_i\hat{S}^z_j$ can be made to vanish for a continuous range of Rabi frequencies with a single microwave frequency; this does not require fine-tuning to a specific point in parameter space.}, Eq.~\eqref{eq:XYZ} becomes a long-ranged version of the XY model in a longitudinal field
\begin{align}
\label{eq:KitaevXY}\hat{H}_{XY}&=\sum_{i,j,i\ne j}\left[J^X_{i,j}\hat{S}^x_i\hat{S}^x_j+J^Y_{i,j}\hat{S}^y_i\hat{S}^y_j\right]-h_z\sum_i \hat{S}^z_i\, .
\end{align}
The nearest-neighbor XY model is equivalent to the Kitaev wire Hamiltonian~\cite{Kitaev}, which has connections to topological phases and Majorana fermions.  It was also pointed out that long-range interactions may not qualitatively change the nature of topological phases~\cite{PhysRevB.87.081106}.  Finally, we note that in the limit of motionally quenched molecules, the quantum statistics of the underlying molecules are unimportant; one can also realize Eq.~\eqref{eq:XYZ} with bosonic or fermionic STMs.

\section{Conclusion}
\label{sec:outlook}

We have identified a general mechanism for generating level crossings between internal states with a finite transition dipole matrix element in symmetric top molecules by a combination of microwave dressing and the linear Stark effect.  Such a pair of near-degenerate dressed states form an effective spin-1/2.  The dipole-dipole interaction generates resonant pair transitions between such nearly degenerate levels.  By appropriate choices of geometry and field polarization, transfer of a single molecule between internal states can be forbidden, and the resulting many-body system features tunable degrees of spatial and spin-component anisotropy.  Using only a single microwave frequency, we show rich tunability of the effective model parameters over a wide range.  As special cases of our general many-body description, we show that Heisenberg XYZ and XY spin models arise when molecules are confined to a one-dimensional line in a deep optical lattice.  Our results provide a new route towards the study of unconventional quantum magnetic phenomena by harnessing the rich internal structure of molecules.

We acknowledge useful conversations with Christina Kraus and Ryan Mishmash during initial development and exploration of the ideas in this work, and thank Kaden Hazzard and Ana Maria Rey for their comments on the manuscript.  This work was supported by {the AFOSR under grants FA9550-11-1-0224 and FA9550-13-1-0086}, {ARO grant number 61841PH, ARO-DARPA-OLE}, and the National Science Foundation under Grants PHY-1207881, PHY-1067973,  PHY-0903457, {PHY-1211914, PHY-1125844}, and NSF PHY11-25915.  We also acknowledge the Golden Energy Computing Organization at the Colorado School of Mines for the use of resources acquired with financial assistance from the National Science Foundation and the National Renewable Energy Laboratories.  We thank the KITP for hospitality.

\emph{Note added:} After this work was submitted, we learned of a related proposal by Glaetzle \emph{et al.} for generating XYZ Heisenberg models using Rydberg atoms (arxiv:1410.3388).  We believe Rydberg atoms offer an exciting alternate scenario for the realization of anisotropic XYZ models, complementary to those described in this paper.

\appendix
\section{Dipole-dipole interactions in microwave-dressed states}
\label{appendix:A}

Here, we consider the long-time averaged matrix elements of the dipole-dipole interaction Eq.~\eqref{eq:compleDDI} in the basis of microwave-dressed states, see Sec.~\ref{sec:DDImicrowave}.  For two states
\begin{align}
|M\rangle&=a|JKM\rangle+be^{-i\omega t}|\left(J+1\right)KM\rangle\, ,\\
|M'\rangle&=c|JKM'\rangle+de^{-i\omega t}|\left(J+1\right)KM'\rangle\, ,
\end{align}
with $M\ne M'$ in the polarization scheme of Fig.~\ref{fig:p=0}(a), we find (see Eqs.~\eqref{eq:dipstart}-\eqref{eq:dipend})
\begin{widetext}
\begin{align}
\label{eq:q01} \langle MM|\hat{H}_{\mathrm{DDI}}|MM\rangle&=\frac{\left(1-3\cos^2\theta\right)}{R^3}\left[\left(a^2D_{J,J;K}^{M,M;0}+b^2D_{J+1,J+1;K}^{M,M;0}\right)^2+2\left(abD_{J,J+1;K}^{M,M;0}\right)^2\right]\, ,\\
\langle MM'|\hat{H}_{\mathrm{DDI}}|MM'\rangle&=\frac{\left(1-3\cos^2\theta\right)}{R^3}\Big[\left(a^2D_{J,J;K}^{M,M;0}+b^2D_{J+1,J+1;K}^{M,M;0}\right)\left(c^2D_{J,J;K}^{M',M';0}+d^2D_{J+1,J+1;K}^{M',M';0}\right)\\
&+2abcdD_{J,J+1;K}^{M,M;0}D_{J,J+1;K}^{M',M';0}\Big]\, ,\\
\label{eq:q02} \langle M'M'|\hat{H}_{\mathrm{DDI}}|M'M'\rangle&=\frac{\left(1-3\cos^2\theta\right)}{R^3}\left[\left(c^2D_{J,J;K}^{M',M';0}+d^2D_{J+1,J+1;K}^{M',M';0}\right)^2+2\left(cdD_{J,J+1;K}^{M',M';0}\right)^2\right]\, ,\\
\label{eq:SPE}\langle MM'|\hat{H}_{\mathrm{DDI}}|M'M\rangle&=\frac{3\cos^2\theta-1}{2R^3}\left[\left(acD_{J,J;K}^{M,M';1}+bdD_{J+1,J+1;K}^{M,M';1}\right)^2+\left(adD_{J,J+1;K}^{M,M';1}\right)^2+\left(bcD_{J+1,J;K}^{M,M';1}\right)^2\right]\, ,\\
\label{eq:q2t}\langle MM|\hat{H}_{\mathrm{DDI}}|M'M'\rangle&=-\frac{3}{2}\frac{\sin^2\theta}{R^3} \Big\{e^{-2i\phi}\left[\left(acD_{J,J;K}^{M,M';1}+bdD_{J+1,J+1;K}^{M,M';1}\right)^2+abcd\left(D_{J,J+1;K}^{M,M';1}\right)^2\right]\, ,\\
&+ e^{2i\phi}\left[\left(acD_{J,J;K}^{M,M';-1}+bdD_{J+1,J+1;K}^{M,M';-1}\right)^2+abcd\left(D_{J,J+1;K}^{M,M';-1}\right)^2\right]\Big\}\, .
\end{align}
\end{widetext}
In addition, terms of the form $\langle M M|\hat{H}_{\mathrm{DDI}}|MM'\rangle$, which cause a transition $|M\rangle\to |M'\rangle$ for one molecule while the other molecule's state is unchanged are also present.  All such terms are proportional to $\sin\theta\cos\theta$ in the present dressing scheme, and so vanish for the 2D geometry of Fig.~\ref{fig:Levels}(b) in which the DC electric field is perpendicular to the plane.  Note that this geometry only refers to the orientation of the DC electric field with respect to the plane, and makes no assumptions about the lattice structure in the plane.  In Eqs.~\eqref{eq:q01}-\eqref{eq:q2t} we explicitly show the $\theta$-dependent factors in to provide clarity about the origins of each term.  In what follows, we assume $\theta=\pi/2$ as in Fig.~\ref{fig:Levels}(b).

To see how these matrix elements can be modified by polarization, let us consider that we have polarization $p=1$ and consider $M'=M+1$, as shown in Fig.~\ref{fig:p=1}.  Here, again neglecting terms which vanish in the 2D geometry of Fig.~\ref{fig:Levels}(b) $(\theta=\pi/2)$, we find
\begin{widetext}
\begin{align}
\nonumber \langle MM|\hat{H}_{DD}|MM\rangle&=\frac{1}{R^3}\left[\left(a^2D_{J,J;K}^{M,M;0}+b^2D_{J+1,J+1;K}^{M+1,M+1;0}\right)^2-\left(abD_{J,J+1;K}^{M,M+1;-1}\right)^2\right]\, ,\\
\langle MM'|\hat{H}_{DD}|MM'\rangle&=\frac{1}{R^3}\Big[\left(a^2D_{J,J;K}^{M,M;0}+b^2D_{J+1,J+1;K}^{M+1,M+1;0}\right)\left(c^2D_{J,J;K}^{M+1,M+1;0}+d^2D_{J+1,J+1;K}^{M+2,M+2;0}\right)\\
\nonumber&-abcdD_{J,J+1;K}^{M,M+1;-1}D_{J,J+1;K}^{M+1,M+2;-1}\Big]\, ,\\
\nonumber \langle M'M'|\hat{H}_{DD}|M'M'\rangle&=\frac{1}{R^3}\left[\left(c^2D_{J,J;K}^{M+1,M+1;0}+d^2D_{J+1,J+1;K}^{M+2,M+2;0}\right)^2-\left(cdD_{J,J+1;K}^{M+1,M+2;-1}\right)^2\right]\, ,\\
\langle MM'|\hat{H}_{DD}|M'M\rangle&=\frac{1}{R^3}\left[\left(bcD_{J+1,J;K}^{M+1,M+1;0}\right)^2-\frac{1}{2}\left(acD_{J,J;K}^{M,M+1;-1}+bdD_{J+1,J+1;K}^{M+1,M+2;-1}\right)^2\right]\, ,\\
\langle MM|\hat{H}_{DD}|M'M'\rangle&=-\frac{3}{2}\frac{1}{R^3} e^{-2i\phi}\left(acD_{J,J;K}^{M,M+1;-1}+bdD_{J+1,J+1;K}^{M+1,M+2;-1}\right)^2\, .
\end{align}
\end{widetext}
Hence, maintaining a single frequency for the microwave field but allowing for different polarizations and intensities realizes extraordinary tunability over the various interaction processes through the coefficients $a$, $b$, $c$, $d$, the dipole matrix elements involved, and the components of the dipole-dipole interaction Eq.~\eqref{eq:q0}-\eqref{eq:q2} which contribute to each process.

\bibliographystyle{prsty}
\bibliography{ST}

\end{document}